\documentstyle[12pt]{article}

\newcommand{\bmat}{\left(\begin{array}}
\newcommand{\emat}{\end{array}\right)}
\def\NPB#1#2#3{Nucl. Phys. B{#1} (19#2) #3}
\def\PLB#1#2#3{Phys. Lett. B{#1} (19#2) #3}

\def\PRD#1#2#3{Phys. Rev. D{#1} (19#2) #3}

\def\MODA#1#2#3{Mod. Phys. Lett.  {#1} (19#2) #3}

\def\yzero{\smash{\hbox{$y\kern-4pt\raise1pt\hbox{${}^\circ$}$}}}

\def\a{\alpha}
\def\b{\beta}
\def\g{\gamma}
\def\d{\delta}
\def\Om{\Omega}

\def\th{\theta}
\def\vt{\vartheta}
\def\vphi{\varphi}
\def\-{\hphantom{-}}
\def\ov{\overline}
\def\s2{\frac{1}{2}}

\def\oh{\frac{1}{2}}
\def\beq{\begin{equation}}
\def\eeq{\end{equation}}
\def\beqa{\begin{eqnarray}}
\def\eeqa{\end{eqnarray}}

\def\Tr{{\rm Tr \,}}
\def\diag{{\rm diag \,}}
\def\IF{\relax{\rm I\kern-.18em F}}
\def\II{\relax{\rm I\kern-.18em I}}
\def\IP{\relax{\rm I\kern-.18em P}}
\def\IC{\relax\hbox{\kern.25em$\inbar\kern-.3em{\rm C}$}}
\def\IR{\relax{\rm I\kern-.18em R}}

\def\cc{{\cal C}}

\def\NN{{\cal N}}

\def\cz{{\cal Z}}

\def\Dsl{\,\raise.15ex\hbox{/}\mkern-13.5mu D} 
\def\IZ{\bf Z}
\def\IC{\bf C}
\def\id{{\rm I}}
\def\idd{{\bf 1}}
\def\Dn{${\ov{{\rm D}9}}$}
\def\Df{${\ov{{\rm D}5}}$}

\newcommand{\drawsquare}[2]{\hbox{%
\rule{#2pt}{#1pt}\hskip-#2pt
\rule{#1pt}{#2pt}\hskip-#1pt
\rule[#1pt]{#1pt}{#2pt}}\rule[#1pt]{#2pt}{#2pt}\hskip-#2pt
\rule{#2pt}{#1pt}}

\newcommand{\fund}{\raisebox{-.5pt}{\drawsquare{6.5}{0.4}}}
\newcommand{\Ysymm}{\raisebox{-.5pt}{\drawsquare{6.5}{0.4}}\hskip-0.4pt%
        \raisebox{-.5pt}{\drawsquare{6.5}{0.4}}}
\newcommand{\Yasymm}{\raisebox{-3.5pt}{\drawsquare{6.5}{0.4}}\hskip-6.9pt%
        \raisebox{3pt}{\drawsquare{6.5}{0.4}}}
\newcommand{\antifund}{\overline{\fund}}
\newcommand{\bYasymm}{\overline{\Yasymm}}
\newcommand{\bYsymm}{\overline{\Ysymm}}

\topmargin
-1.5cm
\textwidth
15.5cm
\textheight
23.5cm
\oddsidemargin
0.7cm
\evensidemargin
1.2cm

\begin{document}
\makeatletter
\@addtoreset{equation}{section}
\makeatother
\renewcommand{\theequation}{\thesection.\arabic{equation}}
\pagestyle{empty}
\rightline{CAB-IB/2911299, IASSNS-HEP-99/79}
\rightline{\tt hep-th/9908072}
\vspace{0.5cm}
\begin{center}
\LARGE{Tachyon-free Non-supersymmetric Type IIB Orientifolds
via Brane-Antibrane Systems \\[10mm]}
\large{G.~Aldazabal$^1$,
A.~M.~Uranga$^2$\\[2mm]}
\small{
$^1$ Instituto Balseiro, CNEA, Centro At\'omico Bariloche,\\[-0.3em]
8400 S.C. de Bariloche, and CONICET, Argentina.\\[1mm]
$^2$ School of Natural Sciences, Institute for Advanced Study \\
Olden Lane, Princeton NJ 08540, USA
\\[9mm]}
\small{\bf Abstract} \\[7mm]
\end{center}
\begin{center}
\begin{minipage}[h]{14.0cm}
We derive the rules to construct type IIB compact orientifolds in
six and four dimensions including D-branes and anti-D-branes. Even
though the models are non-supersymmetric due to the presence of
the anti-D-branes, we show that it is easy to construct large
classes of models free of tachyons. Brane-antibrane annihilation
can be prevented for instance by considering models with branes
and antibranes stuck at different fixed points in the compact
space. We construct several anomaly-free and tachyon-free
six-dimensional orientifolds containing D9-branes and
anti-D5-branes. This setup allows to construct four-dimensional
chiral models with supersymmetry unbroken in the bulk and in some
D-brane sectors, whereas supersymmetry is broken (at the string
scale) in some `hidden' anti-D-brane sector. We present several
explicit models of this kind. We also comment on the role of the
non-cancelled attractive brane-antibrane forces and the
non-vanishing cosmological constant, as providing interesting
dynamics for the geometric moduli and the dilaton, which may
contribute to their stabilization.
\end{minipage}
\end{center}
\newpage
\setcounter{page}{1}
\pagestyle{plain}
\renewcommand{\thefootnote}{\arabic{footnote}}
\setcounter{footnote}{0}

\section{Introduction}

The last two years have witnessed an increased interest on
non-supersymmetric string theory compactifications and non-supersymmetric
states in string theory. This has led to a number of interesting
developments. For instance, the construction of non-supersymmetric string
vacua with vanishing or extremely suppressed cosmological constant
\cite{const}. Another example is the study of tachyon condensation in
brane-antibrane systems and its relation to stable non-BPS states in
string theory \cite{nonbps}. Hopefully these and other developments will
provide new insights into the possible mechanisms breaking supersymmetry
in realistic compactifications of string theory.

One of the most interesting classes of string theory vacua is that
of type IIB orientifolds \cite{sagnotti, dlp,hor,bs,gp,gjdp,afiv}.
Supersymmetric compact orientifolds in six
\cite{bs,gp,gjdp,dp3,bz,kak4} and four
\cite{bl,ang,kak,zwart,afiv} dimensions have been constructed and
studied extensively. They show several interesting properties,
both from the theoretical and the phenomenological points of view.
Within the latter, we may highlight the fact that they provide
explicit string theory models where gauge fields and charged
matter are localized on D-branes, whereas gravity and other closed
string modes propagate in full spacetime. This idea has recently
received a great deal of attention, in particular in the context
of compactifications with large \cite{antoniadis, larged}, intermediate
\cite{bbiq} or `largish' \cite{rs} compact dimensions. It is
natural to wonder about possible supersymmetry breaking mechanisms
in these scenarios.

Several possibilities to achieve the construction of
non-supersymmetric orientifold vacua have been explored in the
literature. For instance, one possibility is to implement the
Scherk-Schwarz mechanism \cite{schsch} in the context of type IIB
orientifolds \cite{schads}. A different approach would consist of
constructing orientifolds of non-supersymmetric string theories, as
in type $0$B orientifolds in \cite{type0}.

In this paper we would like to point out a different possibility.
Non-supersymmetric type IIB orientifolds can be easily obtained by
introducing anti-D-branes in the model. This framework has several
built-in advantages. For example, since the orientifold projections on the
closed string spectrum preserve some of the supersymmetries, closed string
tachyons are automatically absent. Perhaps a bit more surprisingly, open
string tachyons (arising from open strings stretched between branes
and antibranes) are also easily avoided, for instance by choosing
Chan-Paton embeddings that project out the tachyon of coincident branes
and antibranes, or by locating the branes and antibranes stuck at different
fixed points in the internal space. Therefore, large classes of
tachyon-free non-supersymmetric type IIB orientifolds can be constructed
in six and four dimensions. A nice feature that arises in some
four-dimensional models is that they can include chiral $\NN=1$
supersymmetric sectors of D-branes and non-supersymmetric sectors of
anti-D-branes, spatially separated in the compact space. This leads to
explicit string theory models where supersymmetry breaking (at the
string scale) occurs in a distant brane and is transmitted to the visible
supersymmetric sector, either by bulk modes (gravity-mediated models) or
by a gauge sector on other D-branes coupling to both branes and antibranes
(gauge-mediated).

These models also present interesting properties from the
theoretical point of view. For instance we show that certain
non-supersymmetric compactification of type I on orbifold limits
of K3 are related by T-duality to analogous compactification of
the ten-dimensional non-supersymmetric $USp(32)$ theory
constructed in \cite{sugimoto}.

The paper is organized as follows. In Section~2 we provide the basic rules
to construct type IIB orbifold and orientifold models including
antibranes, and compute the spectrum of tachyons and massless states.
This allows to derive conditions under which the open string tachyons
are absent. We also discuss the RR tadpole cancellation conditions for
these models.

In Section~3 we present several examples of non-supersymmetric
type IIB orientifolds in six dimensions. We discuss how some
naively stable models, without tachyons but in which the
antibranes can move off the fixed points into the bulk, can
actually decay to supersymmetric theories through  brane-antibrane
annihilation. We also show how the introduction of antibranes in
some cases allows to satisfy tadpole conditions which cannot be
fulfilled in supersymmetric models with only D-branes. We
illustrate this property by constructing orientifolds of type IIB
on $T^4/\IZ_N$, $N=2,4,6$, with vector structure.

In Section~4, we present four-dimensional examples based on the $\IZ_6'$
and $\IZ_3$ orientifolds, which have unbroken $\NN=1$ supersymmetry in the
closed string sector and in several D-brane sectors, and non-supersymmetric
sectors of anti-D-branes.

Section~5 contains our final remarks and speculations concerning the role
of the non-supersymmetric dynamics in stabilizing the geometric moduli and
the dilaton in these models.

While we were completing the present paper, we noticed reference
\cite{ads}, where one of the models presented in  section 3.1 is
constructed using slightly different techniques.

\section{General construction}

\subsection{Generalities of type IIB orientifolds}

In this section we summarize the basic ingredients
\cite{sagnotti,bs,gp, afiv} and notation needed for the
construction of $N=1$ Type IIB orientifolds \footnote{In the
orientifolds we consider in coming sections, supersymmetry is preserved in
the closed string sector, and is only broken in the open string sector by
the presence of the anti-D-branes. Hence most of the usual techniques
in supersymmetric orientifolds apply.}.

A  type IIB orbifold is obtained when the toroidally compactified
theory is divided out by discrete symmetry group $G_1$, (like
$\IZ_N$ or $\IZ_N\times \IZ_M$).  The orbifold twist eigenvalues $v_a$,
associated to a complex compact coordinate $Y_a$ ($a=1,\dots,
D-10$), are restricted by the requirement that the orbifold group
acts crystalographically on $T^{D-10}$ and by the number of
supersymmetries  to be left unbroken. For instance for $\IZ_N$ in
$D=6$ and  $\NN=2$ unbroken supersymmetries we must have
$v=\frac{1}N(1,-1)$. The same requirement  in $D=4$ is satisfied
by $v={\frac{1}N(\ell_1,\ell_2,\ell_3)}$ with
$\ell_1+\ell_2+\ell_3=0$, where $\ell_a$ are some specific
integers \cite{dhvw}.

A type IIB orientifold results from the joint action of the
orbifold group $G_1$ together with a world sheet parity operation
$\Omega$, exchanging left and right movers. By keeping orientation
reversal invariant states, supersymmetry is reduced by a half.
$\Omega $ action can be also accompanied by extra operations thus
leading to generic orientifold group $G_1+ {\Omega} G_2$ with
${\Omega}h {\Omega} h' \in G_1$ for $h,h' \in G_2$

Orientifolding closed Type IIB string introduces a Klein-bottle
unoriented world-sheet. Amplitudes on such a surface contain
tadpole divergences. Tadpoles may be generically interpreted as
unbalanced orientifold plane charges under RR form potentials.
 In order to eliminate such unphysical divergences Dp-branes, which carry
 opposite charges,  must be generically introduced.
 In this way, divergences occurring in the open string sector cancel  up
the closed sector ones and
produce a consistent theory.

We will focus our discussion in $D=6,4$ dimensional theories.

For $\IZ_N$, with $N$ odd, only D9-branes are required. They fill
the full space-time and the  compact space. For $N$ even,
D5-branes are required. Their  world-volume fills up the
space-time in six dimensions.

$D5_k$-branes, with world volume filling space-time and the
$k^{th}$ complex plane, may be required. This is so whenever the
orientifold group contains the element $\Omega R_i R_j$, for
$k\neq i,j$. Here $R_i$ ($R_j$) is an order two twist of the
$i^{th}$ ($j^{th}$) complex plane.

Open string states are denoted by $|\Psi, ab \rangle $, where
$\Psi$ refers to world-sheet degrees of freedom while the $a,b$
Chan-Paton indices are associated to the open string endpoints
lying on D$p$-branes and D$q$-branes respectively.

These Chan-Paton labels must be contracted with a hermitian matrix
$\lambda ^{pq} _{ab}$.   The action of an element of the
orientifold group on Chan-Paton factors is achieved by a unitary
matrix $\gamma _{g,p}$ such that $g: \lambda ^{pg} \rightarrow
\gamma _{g,p} \lambda ^{pq} \gamma^{-1}_{g,q}$.  We denote by
$\gamma _{k,p}$ the matrix associated to the $Z_N$ orbifold twist
$\theta ^k $ acting on a Dp-brane.

Consistency under group transformations imposes restrictions on
the representations $\gamma _g$. For instance, from $\Omega ^2 =1$
it follows that
\beq \g_{\Om,p }=\pm\g^T_{\Om,p } \label{gomt}
\eeq

Also  group operation   ${(\Om\theta ^k)}^2 =\theta ^{2k}$ and
(\ref{gpa}) lead to
\beq
\g_{k,p}^* =\pm  \g_{\Om,p}^* \, \g_{k,p} \g_{\Om,p}
\label{famp}
\eeq
for $p=9,5$.

Tadpole cancellation imposes further constraints on $\gamma _g$
(see for instance \cite{gjdp,afiv} and references therein).

In what follows we will make a definite choice of signs in
(\ref{gomt}), namely \beqa \g_{\Om,9 } & = & \g^T_{\Om,9 }
\nonumber \\ \g_{\Om,5 } & =& -\g^T_{\Om,5} \label{gpa} \eeqa for
$\Om $ acting on 9 and on 5-branes. The first condition is the
usual requirement of global consistency of the ten-form potential
in Type I theory. Second equation is in agreement with the Gimon
and Polchinski action, analyzed in \cite{gp}.

Generic matrices satisfying above constraints can be provided.
Namely,  for a $Z_N$ orbifold twist action, with $N=2P$ ($N=2P+1$)
we define
\beq
\gamma_{1,p}=({\tilde \gamma_{1,p}},{\tilde \gamma
_{1,p}}^{*})
\eeq
with  $* $ denoting  complex conjugation and
where  ${\tilde \gamma }$ is a $N_p\times N_p$ diagonal matrix
given by
\beq {\tilde \gamma}_{1,p}   = \diag
(\cdots,\alpha^{NV_j}I_{n_j^p},\cdots, \alpha^{NV_ P} I_{n_P^p})
\label{gp}
\eeq

with $\alpha = {\rm e}^{2i\pi /N}$ and $2N_p=2\sum_{j=1}^P  n_j^p$
the number of D$p$-branes.

The choice $V_j=\frac{j}N$ with $j=0,\dots, P$ corresponds to an action
``with vector structure '' ($\gamma ^N=1$) while $V_j=\frac{2j-1}{2N}$
with $j=1,\dots,P$ describes an action ``without vector structure'
($\gamma_{1,p} ^N=-1$) \footnote{Following the classification
introduced in \cite{blpssw} for six-dimensional models.}.

By choosing  $\g_{\Om,9}$ and $\g_{\Om,5}$ matrices
 \beq \g_{\Om,9} =
\bmat{cc}0&\id_{N_9}\\ \id_{N_9} & 0 \emat \quad ; \quad
\g_{\Om,5} = \bmat{cc}0&-i\id_{N_5}\\ i\id_{N_5} & 0 \emat
\label{goms95} \eeq then (\ref{gpa}) and (\ref{famp}) (with $+ (-)$ sign
for D$9(5)$-brane respectively) are satisfied .

The open string spectrum is constructed by requiring the states
$|\Psi, ab \rangle \lambda _{ab}$ to be invariant under the action
of the orientifold group. We present the general rules in the
following section.

\subsection{Open string spectrum with branes and antibranes}

In this section we discuss the rules to construct the spectrum of
tachyonic and massless states in the open string sector, in orientifolds
containing both D-branes and anti-D-branes (denoted ${\ov{\rm D}}$-branes in
what follows). Even though several of the
models we are going to construct are six-dimensional, it will be
convenient to derive the spectrum for a general four-dimensional twist
$v=(v_1,v_2,v_3)=\frac 1N (\ell_1,\ell_2,\ell_3)$.
The rules for six-dimensional models are recovered by setting $v_3=0$ and
$v_1=-v_2=1/N$, and by taking into account that there are two additional
spacetime dimensions.

We start by reviewing the supersymmetric sectors.
\bigskip

\noindent
{\bf The 99 sector}

In the NS sector the GSO projection eliminates the tachyon. Massless gauge
bosons and complex scalars in spacetime are obtained from the states
$\psi_{-\frac 12}^\mu|0,ab\rangle \lambda^{(0)}_{ab}$, $\psi_{-\frac
12}^i|0,ab\rangle \lambda^{(i)}_{ab}$, where $i=1,2,3$ labels the three
complex planes in the orientifold. The projections on the Chan-Paton
factors of these states are
\beqa
\begin{array}{lll}
\lambda^{(0)} = \gamma_{\theta,9} \ \lambda^{(0)} \
\gamma_{\theta,9}^{-1}
& \quad &\lambda^{(0)} = - \gamma_{\Omega,9} \ {\lambda^{(0)}}^T \
\gamma_{\Omega,9}^{-1}\\
\lambda^{(a)} = e^{2\pi i v_a} \ \gamma_{\theta,9}\ \lambda^{(a)}\
\gamma_{\theta,9}^{-1}
& \quad & \lambda^{(a)} = - \gamma_{\Omega,9}\ {\lambda^{(a)}}^T \
\gamma_{\Omega,9}^{-1}
\label{proj1}
\end{array}
\eeqa
In the R sector, states are labeled by weights
$|\pm\s2,\pm\s2,\pm\s2,\pm\s2 \ \rangle$ of a spinor representation of
$SO(8)$, with an odd number of $-\s2$ entries due to the GSO projection.
We obtain  four left-handed spacetime spinors, whose Chan-Paton factors
$\lambda^{(0)}$, $\lambda^{(a)}$ suffer the projections (\ref{proj1}).
Right-handed fermions suffer the opposite projection and provide their
antiparticles. Bosonic and fermionic states form multiplets of $\NN=1$
supersymmetry.
\bigskip

\noindent
{\bf The 55 sector}

The structure of the 55 sector ressembles that of the 99. The only
difference arises in the $\Omega$ projection, which acts with $+1$ sign on
states associated to complex planes with Dirichlet boundary conditions.
The projection for D5$_3$-branes are given by
\beqa
\begin{array}{lll}
\lambda^{(0)} = \gamma_{\theta,5}\ \lambda^{(0)}\ \gamma_{\theta,5}^{-1}
& \quad & \lambda^{(0)} = - \gamma_{\Omega,5}\ {\lambda^{(0)}}^T\
\gamma_{\Omega,5}^{-1}\\
\lambda^{(a)} = e^{2\pi i v_a}\ \gamma_{\theta,5}\ \lambda^{(a)}\
\gamma_{\theta,5}^{-1}
& \quad & \lambda^{(a)} = \pm \gamma_{\Omega,5}\ {\lambda^{(a)}}^T\
\gamma_{\Omega,5}^{-1}
\label{proj2}
\end{array}
\eeqa
with positive sign for $a=1,2$ and negative for $a=3$. The projections
being identical for bosonic and fermionic states, they form $\NN=1$
supermultiplets.
\bigskip

\noindent
{\bf The 59+95 sector}

In the NS sector there are zero modes along the directions with DN
boundary conditions, hence, states are labeled by an internal $SO(4)$ spinor
weight $|s_1,s_2\rangle$, with $s_i=\pm \frac 12$. The GSO projection
requires $s_1=s_2$. The orbifold projection imposes the
following projection on the Chan-Paton factors of 59 and 95 states
\beqa
\lambda_{59} = e^{\pm 2\pi i(v_1+v_2)/2}\ \gamma_{\theta,5}\ \lambda_{59}\
\gamma_{\theta,9}^{-1} & \quad &
\lambda_{95} = e^{\pm 2\pi i(v_1+v_2)/2}\ \gamma_{\theta,9}\ \lambda_{95}\
\gamma_{\theta,5}^{-1}
\eeqa
with the positive and negative signs correspond to the states
$|\s2,\s2\ \rangle$ and $|{\rm -}\s2,{\rm -}\s2\ \rangle$, respectively.
The 59 and 95
sectors are related by the action of $\Omega$ so in fact determining
the spectrum in  one of them is enough.

The zero modes in the R sector arise along the directions with NN boundary
conditions. States are labeled by an $SO(4)$ spinor weight
$|s_3;s_0\rangle$, where $s_0$ determines the spacetime fermion chirality.
The GSO projection requires $s_0=s_3$. The projection for these states is
\beqa
\lambda_{59} = e^{\pm 2\pi iv_3/2}\ \gamma_{\theta,5}\ \lambda_{59}\
\gamma_{\theta,9}^{-1} &\quad &
\lambda_{95} = e^{\pm 2\pi iv_3/2}\ \gamma_{\theta,9}\ \lambda_{95}\
\gamma_{\theta,5}^{-1}
\eeqa
with the positive sign for $|\s2;\s2\ \rangle$ and the negative for
$|{\rm -}\s2;{\rm -}\s2\rangle$. As before, $\Omega$ relates the states in
the 59 and 95 sectors. Notice also that since $\sum_{a=1}^3 v_a=0$, the
states form $\NN=1$ supersymmetric multiplets.

\bigskip

\noindent
{\bf {The ${\bf \bar{9}\bar{9}}$, ${\bf \bar{5}\bar{5}}$,
and ${\bf \bar{5}\bar{9}+\bar{9}\bar{5}}$ sectors.}}

Let us move on to the projection for open string states in
antibrane-antibrane sectors. In these sectors the GSO projection is the
same as in brane-brane sectors. The only difference with respect to the
55, 99 and 59+95 sectors arises in the action of the orientifold
projection on the R states, which has an additional $(-1)$ factor. The
orientifold projection on the NS states remains unchanged.
This fact has been discussed in \cite{sugimoto}, and reflects the fact
that the orientifold projection distinguishes branes from antibranes. In
particular, it respects the  supersymmetries unbroken by the former and
broken by the latter. Notice that this sign flip has no net effect on the
spectrum in the $\bar{5}\bar{9}$, $\bar{9}\bar{5}$ sectors, since they are
not fixed under $\Omega$.

\medskip

Finally, let us discuss the mixed brane-antibrane and antibrane-brane
sectors. As compared with the analogous brane-brane (or
antibrane-antibrane) sectors, they have the opposite GSO projection.
\bigskip

\noindent
{\bf {The ${\bf 9\bar{9}+\bar{9}9}$ sectors}.}

In the NS sector the tachyon $|0,ab\rangle {(\lambda_t)}_{ab}$ survives
the GSO projection, whereas the would-be massless states $\Psi_{-\frac
12}|0\rangle$ do not. The orbifold projection on the Chan-Paton factors
for the tachyons is
\beqa
\lambda_{t,9\bar{9}} = \gamma_{\theta,9}\ \lambda_{t,9\bar{9}}\
\gamma_{\theta,\bar{9}}^{-1} &\quad &
\lambda_{t,\bar{9}9} = \gamma_{\theta,\bar{9}}\ \lambda_{t,\bar{9}9}\
\gamma_{\theta,9}^{-1}
\eeqa
Since $\Omega$ relates the $9\bar{9}$ and $\bar{9}9$ sectors, the tachyons
are real fields. The conditions  under which the
orientifolds are free of tachyons will be discussed below.

In the R sector, the GSO projection selects $SO(8)$ weight vectors $|\pm
\s2,\pm \s2,\pm\s2,\pm\s2\ \rangle$ with an even number of $-\s2$
entries. From these states we get right-handed spacetime fermions
with Chan-Paton factors $\lambda^{(0)}$, $\lambda^{(a)}$ constrained by
the projections
\beqa
\begin{array}{lll}
\lambda^{(0)}_{9\bar{9}}=\gamma_{\theta,9}\ \lambda^{(0)}_{9\bar{9}}\
\gamma_{\theta,\bar{9}}^{-1} & \quad &
\lambda^{(0)}_{\bar{9}9} = \gamma_{\theta,\bar{9}}\
\lambda^{(0)}_{\bar{9}9}\ \gamma_{\theta,9}^{-1} \\
\lambda^{(a)}_{9\bar{9}} = e^{2\pi i v_a}\ \gamma_{\theta,9}\
\lambda^{(a)}_{9\bar{9}}\ \gamma_{\theta,\bar{9}}^{-1} & \quad &
\lambda^{(a)}_{\bar{9}9} = e^{2\pi i v_a}\ \gamma_{\theta,\bar{9}}\
\lambda^{(a)}_{\bar{9}9}\ \gamma_{\theta,9}^{-1}
\end{array}
\eeqa
Actually, $\Omega$ relates states in the $9\bar{9}$ and $\bar{9}9$, so it
is enough to compute only one of these sectors.
\bigskip

\noindent
{\bf {The ${\bf 5\bar{5}+\bar{5}5}$} sector.}

The spectrum in this sector is completely analogous to that in the
$9\bar{9}+\bar{9}9$ sector. The only difference, as we may recall from the
comparison of  55 and 99 sectors, might arise in the projection imposed
by $\Omega$ in directions with D boundary conditions. However, since
$\Omega$ does not impose any projection on either the $5\bar{5}$ or the
$\bar{5}5$ sectors (rather it maps one to the other), no such
difference arises at the level of the spectrum.
\bigskip

\noindent
{\bf {The ${\bf 9\bar{5}+\bar{5}9}$ and ${\bf \bar{9}5+5\bar{9}}$
sectors.}}

These sectors are analogous to the $59+95$ sector, with a few modifications
due only to the opposite GSO projections. The NS states are labeled by a
weight vector $|s_1,s_2\rangle$ with respect to the $SO(4)$ corresponding
to the DN directions. The GSO projection in this case selects $s_1=-s_2$.
The projections on the Chan-Paton factors of these states in the
$9\bar{5}+\bar{5}9$ sector are
\beqa
\lambda_{9\bar{5}} = e^{\pm 2\pi i(v_1-v_2)/2}\ \gamma_{\theta,9}\
\lambda_{9\bar{5}}\ \gamma_{\theta,\bar{5}}^{-1} & \quad &
\lambda_{\bar{5}9} = e^{\pm 2\pi i(v_1-v_2)/2}\ \gamma_{\theta,\bar{5}}\
\lambda_{\bar{5}9}\ \gamma_{\theta,9}^{-1}
\eeqa
with the positive and negative signs for the states $|\s2,
{\rm -}\s2\ \rangle$
and $|\s2,-\s2\rangle$ respectively. The $\Omega$ projection relates the
$9\bar{5}$ and the $\bar{5}9$ sectors. The projections in the
$\bar{9}5+5\bar{9}$ sector are completely analogous.

States in the R sector are labelled by an $SO(4)$ spinor weight
$|s_3;s_0\rangle$, with $s_3=-s_0$ due to the GSO projection, and where
$s_0$ defines the spacetime chirality. The projection on the Chan-Paton
factors in the $9\bar{5}+\bar{5}9$ sector is
\beqa
\lambda_{9\bar{5}} = e^{\pm 2\pi iv_3/2} \gamma_{\theta,9}
\lambda_{9\bar{5}}
\gamma_{\theta,\bar{5}}^{-1} \nonumber \\
\lambda_{\bar{5}9} = e^{\pm 2\pi iv_3/2} \gamma_{\theta,\bar{5}}
\lambda_{\bar{5}9}\gamma_{\theta,9}^{-1}
\eeqa
with the positive and negative signs for the states $|\s2;-\s2\rangle$
and $|-\s2;\s2\rangle$, respectively. As usual $\Omega$ relates the
$9\bar{5}$ and $\bar{5}9$. The projections for the $\bar{9}5+5\bar{9}$
sector are obtained analogously.

\medskip

In order to illustrate the kind of states survive the projections,
let us consider Chan-Paton matrices of the
form
\beqa
\gamma_{\theta,9}=\diag(1_{v_0}, e^{2\pi i\frac 1N} 1_{v_1},\ldots,
e^{2\pi i \frac{N-1}{N}} 1_{v_{N-1}} )
\label{twmat}
\eeqa
and analogous expressions for \Dn-, D5- and \Df-branes,
with the numbers of entries replaced by $w_i$, $n_i$ and $m_i$,
respectively. We must clarify, however, that our projection rules do not
depend on the specific form of these matrices. Indeed, in some of our
models the Chan-Paton matrices will differ from the expression above.

The states resulting from the {\em orbifold} projection with such matrices
are shown in table~1. The subindices denote the gauge group under which
the state transforms. They are defined mod N, hence for negative indices
we have $v_{-i}=v_{N-i}$. The choice of fermions chiralities is mainly the
usual one for supersymmetric sectors. In brane-antibrane sectors, the
opposite chirality arises from the opposite GSO projection. The
orientifold projection will impose additional conditions on these states
in some of the sectors, indicated with an asterisk. For sectors exchanged
by the orientifold projection only one of them is shown. For  these states
the orientifold projection does not impose additional constraints.

\begin{table}[htb]
\small
\renewcommand{\arraystretch}{1.25}
\begin{center}
\begin{tabular}{|c||c|c|c|c|}
\hline
Sector & Gauge bosons & Scalar fields & Fermion$_+$ & Fermion$_-$ \\
\hline\hline
$99^*$ & $\prod_{i=1}^N U(v_i)$ & $\sum_{i=1}^N (v_i,\ov{v}_{i+\ell_a})$
& $\sum_{i=1}^N (v_i,\ov{v}_{i+\ell_a})$ & $\sum_{i=1}^N
{\rm Adj}_{\, i}$ \\
\hline
$\ov{9}\ov{9}^*$ & $\prod_{i=1}^N U(w_i)$ & $\sum_{i=1}^N
(w_i,\ov{w}_{i+\ell_a})$ & $\sum_{i=1}^N (w_i,\ov{w}_{i+\ell_a})$ &
$\sum_{i=1}^N {\rm Adj}_{\, i}$ \\
\hline
$9\ov{9}$ &  & $\sum_{i=1}^N (v_i,\ov{w}_i)$ (tachyons) & $\sum_{i=1}^N
(v_i,\ov{w}_i)$ &
$\sum_{i=1}^N (v_i,\ov{w}_{i+\ell_a})$ \\
\hline
$55^*$ & $\prod_{i=1}^N U(n_i)$ & $\sum_{i=1}^N (n_i,\ov{n}_{i+\ell_a})$
& $\sum_{i=1}^N (n_i,\ov{n}_{i+\ell_a})$ & $\sum_{i=1}^N
{\rm Adj}_{\, i}$ \\
\hline
$\ov{5}\ov{5}^*$ & $\prod_{i=1}^N U(m_i)$ & $\sum_{i=1}^N
(m_i,\ov{m}_{i+\ell_a})$ & $\sum_{i=1}^N (m_i,\ov{m}_{i+\ell_a})$ &
$\sum_{i=1}^N {\rm Adj}_{\, i}$ \\
\hline
$5\ov{5}$ &  & $\sum_{i=1}^N (n_i,\ov{m}_i)$ (tachyons) & $\sum_{i=1}^N
(n_i,\ov{m}_i)$ &
$\sum_{i=1}^N (n_i,\ov{m}_{i+\ell_a})$ \\
\hline
$59$ &  & $\sum_{i=1}^N (n_i,\ov{v}_{i-\frac 12\ell_3})$ & $\sum_{i=1}^N
(n_i,\ov{v}_{i-\frac 12\ell_3})$ & \\
\hline
$\ov{5}\ov{9}$ &  & $\sum_{i=1}^N (m_i,\ov{w}_{i-\frac 12\ell_3})$ &
$\sum_{i=1}^N (m_i,\ov{w}_{i-\frac 12\ell_3})$ & \\
\hline
$5\ov{9}$ & & $\sum_{i=1}^N (n_i,\ov{w}_{i+\frac 12 (\ell_1-\ell_2)})$ &
& $\sum_{i=1}^N (n_i,\ov{w}_{i-\frac 12 \ell_3})$ \\
\hline
$\ov{5}9$ & &  $\sum_{i=1}^N (m_i,\ov{v}_{i+\frac 12 (\ell_1-\ell_2)})$ &
& $\sum_{i=1}^N (m_i,\ov{v}_{i+\frac 12 \ell_3})$ \\
\hline
\end{tabular}
\end{center}
\caption{\small The table shows the spectrum obtained in the different
open string sectors after imposing the orbifold projections, but before
the $\Omega$ projection. To obtain the final spectrum, the orientifold
projection must be imposed on the sectors indicated by an asterisk. In
this last step it must be kept in mind that in antibrane-antibrane sectors
the R states get an additional $(-1)$ factor under the action of $\Omega$.
Scalar fields are complex, save for the tachyon fields which are real.
 \label{specfin} }
\end{table}

One important point to be stressed is that tachyons only
appear when the model contains coincident (or very close) branes of the
same type and with identical Chan-Paton phases. Those tachyons obviously
signal instabilities due to brane-antibrane annihilation. Therefore
the construction of type IIB orientifolds with branes and antibranes but
without tachyons is fairly simple. It only requires that branes and antibranes
sit at different points in the internal space (or have different Wilson
lines, in a T-dual picture), or if they are coincident, that their
Chan-Paton matrices project out the tachyons (that is $n_i=0$ or $m_i=0$,
and $v_i=0$ or $w_i=0$, for all $i=1,\ldots, N$). In Sections~3 and 4 we
give explicit examples of tachyon-free models, and comment on some of
their properties.

\subsection{Cylinder partition function}

In this section we indicate how the open string spectrum derived
above shows up in the cylinder partition functions (see also
\cite{sugimoto,srednicki}). We include it here in order to provide a cross
check for  the correctness of the construction in section 2.1, and  also
in order to illustrate how different signs must be suitable flipped, in a
concrete amplitude, in order to take the presence of anti-branes into
account. We also show how RR tadpole cancellation is achieved by
performing the usual transformation to the closed string channel. We
restrict our analysis to configurations containing D9-branes and
\Dn-branes. Other amplitudes can be similarly analyzed.

The number of states at each mass level $M$ can be obtained by looking
at the multiplicity of the term $q^{M^2}$ ($q=e^{-2\pi t}$) in
the partition function. We will be interested in negative powers,
indicating the presence of a potential tachyon, and in massless fields.

Consider first an array of D$p$ and D$q$-branes. The cylinder amplitudes,
in $D$ dimensions  are given by (see for instance \cite{afiv})
\begin{equation}
\cc_{pq} = \frac{V_D}{2N}\sum_{k=0}^{N-1}
\int_0 ^ \infty \,\frac{dt}t \, (8\pi ^2 \alpha \prime
t)^{\frac{-D}{2}} \cz_{pq}({\theta} ^k) \label{camp}
\end{equation}

where \beq \cz_{pq}(\th^k) =\frac{1}2 \Tr_{pq} \{(1+(-1)^F)  \th^k \, {\rm
e}^{-2\pi t L_0} \} \label{zcdef}
\end{equation}

The trace is over open string states ending at the corresponding brane.
For instance in the ${\bf 99}$  sector, where boundary conditions are NN
in all
directions we have

\beq
\cz_{99}(\th^k) = \frac{1}2\sum_{\a,\b=0,\oh}  \eta_{\a,\b} Z
[{\a \atop \b}]\ (\Tr\g_{k,9}\Tr\g^{-1}_{k,9})
\label{99nsr}
\eeq

with $\eta_{\oh,0}=\eta_{0,\oh}=-\eta_{0,0}=-1$.

Where we have defined

\beq
 Z[{\a \atop \b}]=
{\left[ \frac{\vartheta[{\a \atop \b}]}{\eta^3}\right]^{ \frac{1}{2}(D-2)}\
\prod_{a=1}^{\frac{1}{2}(10-D)}}  \  \frac{\vartheta[{\a \atop {\b
+ k\frac{\ell_a}N}}]}\eta  \, \frac{-2\sin \pi k v_a \,  \eta}
{\vartheta[{\oh \atop {\oh+k\frac{\ell_a}N}}]}.
\label{z991} \eeq

We also define, for later convenience,
\begin{eqnarray}
\cz^{NS}_{99} & =& \frac{1}{2N} \sum_{k=0}^{N-1} [Z [{0 \atop 0}](t)-
Z [{0 \atop \oh}](t)](\Tr\g_{k,9}\Tr\g^{-1}_{k,9}) \nonumber \\
\cz^{R}_{99} & =& -\frac{1}{2N} \sum_{k=0}^{N-1} Z [{\oh \atop
0}](t)(\Tr\g_{k,9}\Tr\g^{-1}_{k,9})
\label{z99}
\end{eqnarray}

which are the separate contributions of open string NS bosons and R
fermions respectively.  Recall also that $Z [{0 \atop \oh}](t)$
correspond to RR fields in the closed string sector.

In order to extract the leading behaviour for $q=e^{-2\pi t}$ we
use the product form of $\vartheta$ function
\beq
\frac{\vt[{\d \atop \vphi}]}{\eta} = {\rm e}^{2i\pi \d \vphi}
\, q^{\oh \d^2 - \frac1{24}} \, \prod_{n=1}^\infty (1 + q^{n+\d
-\oh} {\rm e}^{2i\pi \vphi} ) \, (1 + q^{n-\d -\oh} {\rm
e}^{-2i\pi \vphi} )
\label{vtp}
\eeq
where the Dedekind $\eta$ function is \beq \eta =  q^{\frac1{24}} \,
\prod_{n=1}^\infty (1 -q^n)
\label{deta}
\eeq

Thus, collecting all terms we finally have, for $\eta_{\a,\b}
Z[{\a \atop \b}]$

\begin{eqnarray}
Z [{0 \atop 0}](t) &=& q^{-\oh} +
[D-2+ 2\sum_{a=1}^{\frac{1}{2}(10-D)} \cos(\frac{2\pi k \ell_a}{N})] +
 O(q)+\dots \\
-Z [{0 \atop \oh}](t) &=& -q^{-\oh}
+[D-2+2\sum_{a=1}^{\frac{1}{2}(10-D)} \cos(\frac{2\pi k\ell_a}{N})]
+ O(q)+\dots \nonumber \\
-Z [{\oh \atop 0}](t) &=& -16\prod_{a=1}^{\frac{1}{2}(10-D)}
\cos(\frac{\pi k\ell_a}{N})
\\
-Z [{\oh \atop \oh}](t)  &=&0
\label{zexp}
\end{eqnarray}
We see that the potential  tachyon field associated to  $q^{-\oh}$
disappears from the spectrum. Also, since $D-2+ 2\sum_{a=1}^{\frac{1}{2}
(10-D)} \cos(\frac{2\pi k\ell_a}{N})=8\prod_{a=1}^{\frac{1}{2}(10-D)}
\cos(\frac{\pi k\ell_a}{N})$   ( for
$\sum_{a=1}^{\frac{1}{2}(10-D)}\ell_a=0$)
bosons and fermions exactly cancel in $\cz_{99}$ at this order (and to
all orders due to Riemann identities \cite{mum}), exhibiting supersymmetry.

The traces of the twist matrices defined in (\ref{twmat}) are
\begin{equation}
\Tr\g_{k,9}= \sum_{j=0}^{N-1} e^{\frac{2i\pi kj}N}v_j
\label{t9}
\end{equation}
and similarly for $\Tr\g^{-1}_{k,9}$. Hence, by introducing these
expressions in (\ref{z99}) and by performing the sum over $k$ we
finally obtain
\begin{equation}
\cz ^{NS}_{99}= \sum_{i=0}^{N-1} \sum_{a=1}^{\frac{1}{2}(10-D)}
[(D-2)v_iv_i+ v_iv_{i+\ell_a}+ v_{i+\ell_a}v_{i}]+O(q)+\dots
\end{equation}

with the convention $v_i=v_{N+i}$. These are the exact
multiplicities for the gauge bosons $\prod_{i=1}^N U(v_i)$ (the factor
$D-2$ corresponding  to the number of transverse polarizations), and the
scalars computed in section 2.2.
Recall that an extra $\oh$ factor must be included when the
orientifold projection is performed. Thus a multiplicity
$\oh v_i^2$ would appear. This completes to $\oh v_i(v_i-1)$
multiplicity of $SO(v_i)$ group when the unoriented string amplitudes are
included.

For fermions we have $\cz ^{R}_{99}=-\cz ^{NS}_{99}$. Multiplicities  are
now interpreted as coming from positive (internal) chirality fermions and
negative chirality adjoint fermions as shown in Table 1.

Multiplicities in {\bf {$\ov 9 \ov 9$}} sector are obtained by
simply replacing $v_i \rightarrow w_i$ above.

In order to obtain the states multiplicities in   {\bf {$\ov 9  9 +  9
\ov 9$}} cylinders we must consistently change the sign of the closed
sector RR charges. This amounts to flipping the sign of  $Z [{0 \atop
\oh}](t)$ above. Namely
\begin{eqnarray}
\cz^{NS}_{9\ov 9+\ov 9 9} & =& \frac{1}{2N}
\sum_{k=0}^{N-1} [Z [{0 \atop 0}](t)+Z [{0 \atop \oh}](t)]
(\Tr\g_{k,\ov 9}\Tr\g^{-1}_{k,9}+ \Tr\g_{k, 9}\Tr\g^{-1}_{k,\ov 9})
\nonumber \\
\cz^{R}_{9\ov 9+\ov 9 9} & =& -\frac{1}{2N}
\sum_{k=0}^{N-1} [Z [{\oh \atop 0}](t)]
(\Tr\g_{k,\ov 9}\Tr\g^{-1}_{k,9}+\Tr\g_{k, 9}\Tr\g^{-1}_{k,\ov 9})
\label{z99b}
\end{eqnarray}

Hence, we notice, by recalling the expansions (\ref{zexp}), that NS
tadpoles do not cancel. In fact, we find

\begin{eqnarray}
\cz^{NS}_{9\ov 9+\ov 9 9} & =& q^{-\oh} \sum_{j=0}^{N-1} 2v_jw_j +
O(q)+\dots
\label{z99bNS}
\end{eqnarray}
which  corresponds to  {\bf {${\ov 9} 9$}} sector tachyons
$\sum_{i=0}^{N-1}[(v_i,\ov w_i)+(\ov v_i, w_i)]$.

Finally for fermions we obtain
\begin{equation}
\cz^{R}_{9\ov9 +\ov9 9}= 2\ 2^{\frac{D-4}{2}}[\sum_{i=0}^{N-1}
2v_iw_i +  \sum_{a=1}^{\frac{1}{2}(10-D)}
v_iw_{i+\ell_a}+ v_{i+\ell_a}w_{i}]+O(q)+\dots
\label{z99bR} \end{equation}
Again, these are the multiplicities which correspond to
massless fermionic states obtained above.
\bigskip

\noindent
{\bf RR tadpole cancellation}
\bigskip

Cancellation of RR tadpoles is a basic requirement for the consistency of
the theory. In Type IIB theory with only D-branes (no anti-branes),
tadpole cancellation is equivalent to anomaly cancellation in $D=10,6$
dimensions, however, it is generally stronger in $D=4$ \cite{abiu}. Also,
if RR tadpoles are absent so must be NS tadpoles due to supersymmetry.
This last result is thus not expected in models with anti-branes where
supersymmetry is broken. NS tadpoles should manifest as a background
redefinition \cite{fs}.

In order to analyze the tadpole divergences it is useful to rewrite the
partition function as
\begin{eqnarray}
\cz^{NS} &= &-\frac{1}{2N} \sum_{k=0}^{N-1} \left\{ Z [{0 \atop \oh }](t)
(\Tr\g_{k,9}-\Tr\g_{k,\ov 9})(\Tr\g^{-1}_{k,9}-\Tr\g^{-1}_{k,\ov 9})+
\right. \nonumber\\
& & \left. Z [{0 \atop 0}](t)
(\Tr\g_{k,9}+\Tr\g_{k,\ov 9})(\Tr\g^{-1}_{k,9}+\Tr\g^{-1}_{k,\ov 9})
\right\}
\label{zns}
\end{eqnarray}
and
\begin{equation}
\cz^{R}= - \frac{1}{2N}
\sum_{k=0}^{N-1} Z [{\oh \atop 0}](t)
(\Tr\g_{k,9}+\Tr\g_{k,\ov 9})(\Tr\g^{-1}_{k,9}+\Tr\g^{-1}_{k,\ov 9})
\end{equation}

where we have used (\ref{z99}) (and a similar expression with
$9\rightarrow \ov 9$ for the $\ov 9\ov 9$ sector) and (\ref{z99b}).

RR tadpoles divergences are contained in $Z [{0 \atop \oh }](t)$ and have
been analyzed in different situations in Type IIB. We rederive them here
for completeness.  In fact, by performing a modular transformation
$t \rightarrow 1/t $ and looking at small values of $t$ it can
easily
obtain
\beq
Z [{0 \atop \oh}](1/t)=
t^{\frac{1}{2}(D-2)}\prod_{a=1}^{\frac{1}{2}(10-D)} (-2\sin
\frac{\pi k\ell_a}{N})
\label{rrt}
\eeq
Nevertheless we note that, regarding tadpole cancellation, the net effect
of the introduction of anti D9-branes, with respect to Type IIB theory
with only D9-branes, is that we must replace
\begin{equation}
\Tr\g_{k, 9}\rightarrow \Tr\g_{k, 9}-\Tr\g_{k,\ov 9}
\label{tc}
\end{equation}
since we must require the coefficient of $Z [{0 \atop \oh}]$ in
(\ref{zns}) to vanish. The rule (\ref{tc}) reflects the fact that branes
and antibranes have opposite charges with respect to the RR fields.

For instance, in a six-dimensional type IIB orbifold compactification
(non orientifold yet), the only tadpoles arise from cylinder
amplitudes. The corresponding equations can be extracted from \cite{gjdp}
and lead to
\begin{equation}
\Tr\g_{k, 9}-\Tr\g_{k,\ov 9}-4\sin^2\frac{\pi k}N \,(\,\Tr\g_{k,
5,L}-\Tr\g_{k,\ov 5,L}\, )\,=0
\end{equation}
for $5,\ov 5$-branes at orbifold fixed point $L$.

Notice that by computing the traces as in (\ref{t9}), these equations are
equivalent to (we have dropped the $L$ index for $m_i,n_i$ here)
\begin{equation}
-2n_i+n_{i+1}+n_{i-1}+v_i+2m_i-m_{i+1}-m_{i-1}=0
\end{equation}
Precisely, these are the conditions for cancellation of anomalies of the
six dimensional gauge theories living on branes and anti-branes with the
spectrum computed in section 2.1 (before the orientifold projection).

It is also possible to show that the Green-Schwarz mechanism that cancels
the residual anomalies in six \cite{sagnan,blpssw} and four \cite{iru1}
dimensions works in models with antibranes. In this respect, it is
important to notice that the fermions in brane-antibrane sectors have
chirality opposite to that of usual matter fermions in supersymmetric
models. Hence their contribution to the anomaly has the opposite sign as
well. This sign also emerges in the GS counterterms because the coupling of
antibranes to RR fields is opposite to that of branes. Hence cancellation
follows in the usual way. For antibrane-antibrane sectors, fermions
have the usual chirality, and the GS countertems also have the usual sign
(since the $-1$ signs of the two antibrane couplings give an overall
$+1$).

\section{Some six-dimensional examples}

In this Section we construct several explicit non-supersymmetric type IIB
orientifolds in six dimensions with chiral spectrum. These models are free
of tachyons, and illustrate a number of interesting generic features of
compactifications with branes and antibranes. Also, since chiral theories
are potentially anomalous in six dimensions, the cancellation of anomalies
in the models we present serves as a useful check of the rules proposed
to construct the spectrum in Section~2.

\subsection{A non-supersymmetric $Z_3$ model}

A simple possibility to obtain non-supersymmetric models is to modify
slightly one of the familiar supersymmetric compactifications by the
introduction of antibranes, in a way consistent with cancellation of RR
tadpoles. In order to illustrate the basic idea, let us consider a simple
model related to the six-dimensional $\IZ_3$ model in \cite{gjdp}. The
construction will be mostly intuitive in terms of a T-dual version, where
$T^4/\IZ_3$ is modded out by $\Omega R_1R_2$, and the model contains no
D9-branes. It contains D5-branes (and \Df-branes) sitting at
points in the compact space. As discussed in Section~2.2, the twisted
tadpole cancellation conditions for the branes at the origin are obtained
from those in \cite{gjdp} by replacing $\Tr \gamma_{\theta^k,5}\to \Tr
\gamma_{\theta^k,5}-\Tr\gamma_{\theta^k\bar{5}}$, so we have
\beqa
& \Tr\gamma_{\theta,5}-\Tr\gamma_{\theta,\bar{5}}=8 & \nonumber \\
&\Tr\gamma_{\theta^2,5}-\Tr\gamma_{\theta^2,\bar{5}}=8 &
\eeqa
These can be satisfied, for instance, by choosing
\beqa
\gamma_{\theta,5}=\idd_4 \quad ; \quad
\gamma_{\theta,\bar{5}}=\diag(e^{2\pi
i\frac 13}1_4,e^{2\pi i\frac 23}1_4)
\eeqa
The matrices $\gamma_{\Omega R_1R_2}$ simply exchanges opposite phases in
$\gamma_{\theta}$. The untwisted tadpole requires a net D5-brane number of
32, $n_5-n_{\bar{5}}=32$, and so the model must contain additional
D5-branes.
Let us consider placing 18 D5-branes at one of the orbifold
(rather than orientifold) points fixed under $\IZ_3$, and another set of
18 at its image under $\Omega R_1R_2$. For an orbifold point, twisted
tadpole conditions amount to tracelessness of the relevant Chan-Paton
matrices. Therefore these D5-branes are described by
\beqa
\gamma_{\theta,5}=\diag(1_6,e^{2\pi i\frac 13}1_6,e^{2\pi i\frac 23}1_6)
\eeqa
The gauge group arising from branes (and antibranes) at the origin is
$SO(4)_{55}\times U(4)_{\bar{5}\bar{5}}$, with non-supersymmetric matter
content given by \footnote{Notice that the projection with $\Omega R_1R_2$
has some additional $(-1)$ signs as compared with the $\Omega$ projection.}
\beqa
\begin{array}{cccc}
{\rm {\bf Sector}} & {\rm {\bf Complex\;\, Scalars}} & {\rm{\bf
Fermion}}_+ & {\rm {\bf Fermion}}_-\\
55 & & & ({\rm Adj},1) \\
\bar{5}\bar{5} & (1,\Yasymm)+(1,\bYasymm) & (1,\Ysymm) & (1,{\rm Adj}) \\
5\bar{5}+\bar{5}5 & & & (\fund,\fund)
\end{array}
\eeqa
The D5-branes sitting at the orbifold fixed points give a $D=6$, $\NN=1$
spectrum with group $U(6)^3$, and hypermultiplets in $(\fund,\antifund,1)+
(1,\fund,\antifund)+(\antifund,1,\fund)$.

Since the orientifold projection on the closed string sector is exactly
that in \cite{gjdp}, it gives rise to the $D=6$, $\NN=1$ supergravity and
dilaton multiplets, and the following set of $D=6$, $\NN=1$ matter
multiplets: two hypermultiplets in the untwisted sector, and nine hyper-
and nine tensor multiplets in the twisted sectors. It is easy to check
that both gauge and gravitational irreducible anomalies cancel in the
model. As usual, the residual non-abelian and $U(1)$ anomalies are
cancelled by the GS mechanism mediated by the tensor and
hypermultiplets \cite{sagnan,blpssw}.

This theory can be thought of as a toy model where we have a supersymmetric
sector of branes and a non-supersymmetric sector containing antibranes.
Supersymmetry breaking would be transmitted from the latter to the former
by exchange of bulk fields. However, the model suffers from the following
serious drawback. As discussed in a number of papers (see {\em e.g.}
\cite{banksussk}), the lack of BPS properties in brane-antibrane
systems leads to an attractive force between the two kinds of objects.
In brane-brane (or antibrane-antibrane) systems, the repulsive force
mediated by exchange of RR closed string fields is exactly cancelled by
that from exchange of NS-NS fields, leading to no net force. In
brane-antibranes systems, however, the RR piece becomes attractive and
branes and antibranes tend to come close to each other. Once the distance
reaches a critical value, a tachyon develops, signaling the possibility of
annihilation and decay to the vacuum. This type of phenomenon takes place
in the model we have constructed. Indeed, the D5-branes sitting
at the $\IZ_3$ orbifold point are not stuck to it, and are free to move
into the bulk as a dynamical D5-brane. This transition very likely takes
place due to the attractive force induced by the D5-branes located (and
actually, stuck) at the origin. Once in the bulk, the D5-branes approach
the origin, tachyons develop, branes and antibranes annihilate and the
theory decays to the supersymmetric model in \cite{gjdp}, where the
Chan-Paton matrices for D5-branes are
\beqa
\gamma_{\theta,5}=\diag(1_{16}, e^{2\pi i\frac 13} 1_{8}, e^{2\pi i\frac
23} 1_8)
\eeqa

Therefore, the initial mode should be thought of as an excited
non-supersymmetric state of this supersymmetric vacuum. It is
straightforward to construct this type of non-supersymmetric
compactifications, by nucleating brane-antibrane pairs in any
supersymmetric model, and separating the members of these pairs. Since
this type of models is not really new or interesting for our purposes, the
models we construct in the following will be safe against this type of
decay. This can be achieved, as we show in Section~4  by considering
branes and antibranes stuck at different fixed points. A  different
possibility is to consider branes and antibranes with different
world-volume dimension (since tachyons only appear in $p{\ov p}$ sectors).
Models of this kind are presented in next subsection.

\subsection{Non-supersymmetric models with vector structure}

In this section we construct a different kind of models, quite unrelated
to supersymmetric orientifolds. They illustrate the fact
that the introduction of antibranes sometimes allows to satisfy RR tadpole
cancellation conditions which could not be satisfied in supersymmetric
models containing only D-branes.

The models we consider are obtained by modding out type IIB theory on
$T^4/\IZ_N$ (with $N=2,4,6$)  by an orientifold projection $\Omega$ that
preserves $D=6$, $\NN=1$ supersymmetry in the closed string spectrum. The
orientifold projection we are interested in discussing, however, differs
from that in \cite{gjdp} and was first discussed in \cite{polchinski}.
Concretely, its action on the order two twisted sector of $\IZ_N$ is such
that the RR states are left-right symmetric (rather than antisymmetric)
combinations. In other words, the orientifold projection gives rise to a
tensor multiplet (rather than a hypermultiplet) in the $\IZ_2$ twisted
sectors. This projection imposes a constraint on the Chan-Paton matrices
of the corresponding twist
\cite{polchinski}
\beqa
\gamma_{\Omega}=\gamma_{\theta^{N/2}} \gamma_{\Omega}
\gamma_{\theta^{N/2}}^T
\eeqa
for D9- and D5-branes (and also for antibranes). The constraint implies
the bundles on the D-branes
have vector structure (see \cite{blpssw} for a discussion of vector
structure in orientifold models). This type of projection was studied in
\cite{bi,pu} in the context of orientifolds of $\IC^2/\IZ_N$, which
provide a local description of the fixed point of the compact models.

It is not possible to construct a supersymmetric orientifold of
$T^4/\IZ_N$ where all fixed points are of this type \footnote{It is
however possible to construct mixed models with fixed points of different
kinds, see \cite{polchinski} for a $\IZ_2$ example.}. The reason is that
the cancellation of the untwisted tadpole generated by the orientifold
planes requires $n_5=-32$. Clearly it is not possible to cancel this tadpole
by introducing D5-branes. However, as we show below, it is certainly
possible to construct consistent tachyon-free non-supersymmetric models,
where this tadpole is neatly cancelled by the introduction of $32$
\Df-branes. Untwisted tadpoles also require $n_9=32$, as in the
usual case. In these models, tachyons are absent because branes and
antibranes have different world-volume dimension, hence no anihilation is
possible.

Before entering the detailed construction of the model, let us introduce
some conventions, which differ slightly from those in section 2.1. The
Chan-Paton matrices for the twist $\gamma_{\theta}$ will be of the form
\beqa
\gamma_{\theta, 9}=\diag(1_{v_0},e^{2\pi i\frac 1N} 1_{v_1},\ldots,
e^{2\pi i\frac{N-1}{N}} 1_{v_{N-1}})
\eeqa
(and analogously for $\gamma_{\theta,\bar{5}}$), with $v_i=v_{N-i}$ due
to the orientifold symmetry. The matrices $\gamma_{\Omega,9}$ will be of
the form
{\footnotesize
\beqa
\gamma_{\Omega,9}=\pmatrix{
1_{v_0} & & &  & & \cr
        & & & &  & 1_{v_1} \cr
        & & & & \ldots & \cr
        & & & 1_{v_{N/2}}& & \cr
        & & \ldots& & & \cr
        & 1_{v_{N-1}} & &  & & \cr
} \quad ;\quad \gamma_{\Omega,\bar{5}}=\pmatrix{
\epsilon_{v_0} & & &  & & \cr
        & & & &  & 1_{v_1} \cr
        & & & & \ldots & \cr
        & & & \epsilon_{v_{N/2}}& & \cr
        & & \ldots& & & \cr
        & -1_{v_{N-1}} & &  & & \cr
}\nonumber
\eeqa
}
These conventions are useful to make contact with \cite{bi,pu}, where the
twisted tadpoles for these orientifolds were computed. The general
expression can be extracted from these references \footnote{The tadpole
conditions in \cite{bi} are slightly different due to a different form of
the matrices $\gamma_{\theta}$. Our expression is taken from
section 3.2 in \cite{pu}, with the only modifications of multiplying the
crosscap by a factor of $4$ (since our models are six-dimensional rather
than four-dimensional) and replacing $\Tr\gamma_{\theta^k,5}$ by $-\Tr
\gamma_{\theta^k,\bar{5}}$.}
\beqa
\Tr \gamma_{\theta^k,9}+4\sin^2 \frac{\pi k}{N}\Tr
\gamma_{\theta^k,\bar{5}} -32 \delta_{k,0\;{\rm mod} \; 2} = 0
\label{tadpovec}
\eeqa
for branes at $\IZ_N$ fixed points. Since these models are in a sense the
simplest six-dimensional compactifications with vector structure, we now
turn to the explicit construction of the $\IZ_2$, $\IZ_4$ and $\IZ_6$
models.

\subsubsection{The $\IZ_2$ model}

Consider modding out Type IIB on $T^4/\IZ_2$ by the orientifold
projection $\Omega$ that selects the tensor multiplet in all $\IZ_2$
twisted sectors. The closed string spectrum gives a set of $D=6$,
$\NN=1$ multiplets, concretely 4 hypermultiplets in the untwisted sector
and 16 tensor multiplets in the twisted sector.

Concerning the open string spectrum, we consider, as discussed above,
$n_9=n_{\bar{5}}=32$ to satisfy the untwisted tadpoles. We will also
consider the model where all the \Df-branes sit at one $\IZ_2$
fixed point. Other models can be constructed analogously. RR tadpoles for
such configuration are cancelled by choosing
\beqa
\gamma_{\theta,9}&=&\diag(1_{16},-1_{16}) \nonumber \\
\gamma_{\theta,\bar{5}}&=& \diag(1_{16},-1_{16})
\eeqa
The gauge group is $[SO(16)\times SO(16)]_{99}\times [USp(16)\times
USp(16)]_{\bar{5}\bar{5}}$. The matter content arising from the
projections in Section~2 is
\beqa
\begin{array}{llc}
{\bf 99} \; {\rm sector :} & {\rm Complex \; Scalars:} & 2(16,16) \\
                     & {\rm Fermions}_+\; : & (16,16) \\
                     & {\rm Fermions}_- \; :& (120,1)+(1,120) \\
{\bf \bar{5}\bar{5}} \; {\rm sector :} & {\rm Complex \; Scalars:} &
2(16,16) \\
                     & {\rm Fermions}_+ \; :& (16,16) \\
                     & {\rm Fermions}_- \; :& (120,1)+(1,120) \\
{\bf \bar{5}9+9\bar{5}} \; {\rm sector :} &  {\rm Complex \; Scalars:} &
(16,1;1,16) + (1,16;16,1) \\
 & {\rm Fermions}_- \; :& \frac 12 (16,1;16,1) + \frac 12 (1,16;1,16)
\end{array}
\eeqa
Here the ${\bf 120}$ of $USp(16)$ is actually reducible as ${\bf 119} +
{\bf 1}$. Also, the factor $\frac 12$ in front of some
fermion representations implies they have a (symplectic) Majorana-Weyl
constraint, which is possible because they transform in pseudoreal
representations of the gauge group (these are the fermions appearing in
the more familiar half-hypermultiplets of $D=6$, $\NN=1$ supersymmetric
models).

It is easy to check that all irreducible gauge and gravitational anomalies
cancel in the model above. As usual, we expect the residual non-abelian
and $U(1)$ anomalies to be cancelled through a Green-Schwarz mechanism
\cite{sagnan}.

\subsubsection{The $\IZ_4$ model}

Let us discuss the construction of a orientifold of $T^4/\IZ_4$ by this
orientifold projection. In the closed string sector we obtain a set of
$D=6$, $\NN=1$ multiplets. Specifically, the untwisted sector provides
two hypermultiplets. Concerning the twisted sector, the model contains
four $\IC^2/\IZ_4$ fixed points, each of which contributes with one hyper-
and two tensor multiplets, and six $\IC^2/\IZ_2$, each contributing with
one tensor multiplet. Hence we get 6 hyper- and 14 tensor multiplets.

Regarding the open string sector, untwisted tadpole cancellation
conditions require $n_9=n_{\bar{5}}=32$. It is possible to show that in
the $\IZ_4$ model tadpole cancellation conditions do not have solutions if
all \Df-branes are located at a single fixed point. However, if we
place \Df-branes at all four $\IZ_4$ fixed points, it is possible
to solve the tadpole conditions. In what follows we describe a simple
solution, even though there are other possibilities.

Let us consider the D9-branes to be described by the following Chan-Paton
matrix
\beqa
\gamma_{\theta,9}=\diag(1_8,e^{2\pi i\frac 14} 1_8, e^{2\pi i\frac 24}
1_8, e^{2\pi i\frac 34} 1_8)
\eeqa
We also consider a set of 8 \Df-branes sitting at each of the four
$\IZ_4$ fixed points labeled by $L=1,\ldots, 4$. Their Chan-Paton matrix is
\beqa
\gamma_{\theta,\bar{5}, L}=\diag(1_4,e^{2\pi i\frac 24} 1_4)
\eeqa
These choices satisfy the tadpole conditions (\ref{tadpovec}) at each
$\IZ_4$ fixed point.

The gauge group in this model is
\beqa
SO(8) \times U(8) \times SO(8) \times \prod_{L=1}^4 [USp(4)\times
USp(4)]_L
\eeqa
The matter content is given by
\beqa
\begin{array}{llc}
{\bf 99} \; {\rm sector :} & {\rm Complex \; Scalars:} & (8,\bar{8},1) +
(1,8,8) + (8,8,1) + (1,\bar{8},8)\\
                     & {\rm Fermions}_+ \; :& (8,\bar{8},1) + (1,8,8) \\
                     & {\rm Fermions}_- \; :& (28,1,1) + (1,64,1)
+(1,1,28) \\
{\bf \bar{5}_L\bar{5}_L} \; {\rm sector :} & {\rm Fermions}_- &
(6_L,1)+(1,6_L)
\\
{\bf \bar{5}_L9+9\bar{5}_L} \; {\rm sector :} &  {\rm Complex \; Scalars:}
& (4_L,1;1,\bar{8},1) + (1,4_L;1,8,1)\\
 & {\rm Fermions}_- \; :& \frac 12 (4_L,1;8,1,1) + \frac 12 (1,4_L;1,1,8)
\end{array}
\eeqa
where the ${\bf 6}$ of $USp(4)$ is actually reducible as ${\bf 5} + {\bf
1}$. It is easy to see that the complete open and closed string spectrum
cancels all gauge and gravitational anomalies.

\subsubsection{The $\IZ_6$ model}

Let us consider the six-dimensional $\IZ_6$ model, with the orientifold
projections selecting the tensor multiplets in $\IZ_2$
twisted sectors. The closed string sector gives rise to a set of $D=6$,
$\NN=1$ multiplets. There are two hypermultiplets arising from
the untwisted sector. The model contains one $\IC^2/\IZ_6$ fixed points,
which contributes two hyper- and three tensor multiplets, four
$\IC^2/\IZ_3$ orbifold points, contributing one hyper- and one tensor
multiplet each, and five $\IC^2/\IZ_2$ points, contributing one tensor
multiplet each. We have a total of 8 hyper- and 12 tensor multiplets.

Concerning the open string spectrum, we have as usual $n_9=n_{\bar{5}}=32$.
In this case there are solutions to the twisted tadpole cancellation
conditions with all \Df-branes sitting at the origin. The general
solution for the Chan-Paton matrices in such configuration is
\beqa
\gamma_{\theta,9}=\diag(1_{2K}, e^{2\pi i\frac 16} 1_{K}, e^{2\pi
i\frac 26} 1_{8-K}, e^{2\pi i\frac 36} 1_{16-2K}, e^{2\pi i\frac 46}
1_{8-K}, e^{2\pi i\frac 56} 1_{K} ) \nonumber \\
\gamma_{\theta,\bar{5}}=\diag(1_{16-2K}, e^{2\pi i\frac 16} 1_{8-K},
e^{2\pi i\frac 26} 1_{K}, e^{2\pi i\frac 36} 1_{2K}, e^{2\pi i\frac 46}
1_{K}, e^{2\pi i\frac 56} 1_{8-K} )
\eeqa
A particularly nice solution is obtained for $K=4$, on which we center for
the sake of concreteness, the spectrum for general $K$ being analogous.
The gauge group in such case is
\beqa
SO(8)\times U(4)\times U(4)\times SO(8)\times USp(8) \times U(4) \times
U(4) \times USp(8)
\eeqa
The matter spectrum is given by
{\small
\beqa
\begin{array}{llc}
{\bf 99} \; {\rm sector :} & {\rm Complex \; Scalars:} & (8,\ov{4},1,1) +
(1,4,\ov{4},1) + (1,1,4,8) + \\
 & & + (8,4,1,1) + (1,\ov{4},4,1) + (1,1,\ov{4},8) \\
                     & {\rm Fermions}_+ \; :& (8,\ov{4},1,1) +
(1,4,\ov{4},1) + (1,1,4,8) \\
                     & {\rm Fermions}_- \; :&
(28,1,1,1) + (1,16,1,1) + \\
& & +(1,1,16,1) + (1,1,1,28) \\
{\bf \ov{5}\ov{5}} \; {\rm sector :} & {\rm Complex \; Scalars:}
& (8,\ov{4},1,1) + (1,4,\ov{4},1) + (1,1,4,8) + \\
 & & + (8,4,1,1) + (1,\ov{4},4,1) + (1,1,\ov{4},8) \\
                     & {\rm Fermions}_+\; : & (8,\ov{4},1,1) +
(1,4,\ov{4},1) + (1,1,4,8) \\
                     & {\rm Fermions}_- \; :&
(28,1,1,1) + (1,16,1,1) + \\
& & +(1,1,16,1) + (1,1,1,28) \\
{\bf \ov{5}9+9\ov{5}} \; {\rm sector :} &  {\rm Complex \; Scalars:} &
(8,1,1,1;1,\ov{4},1,1) + (1,4,1,1;1,1,\ov{4},1) + \\
& & + (1,1,4,1;1,1,1,8)+ (1,\ov{4},1,1;8,1,1,1) + \\
& &+(1,1,\ov{4},1;1,4,1,1) + (1,1,1,8;1,1,4,1) \\
 & {\rm Fermions}_- \; :& \frac 12(8,1,1,1;8,1,1,1) +
(1,4,1,1;1,\ov{4},1,1) + \\
& & + (1,1,4,1;1,1,\ov{4},1) + \frac 12 (1,1,1,8;1,1,1,8)
\end{array} \nonumber
\eeqa}
The ${\bf 28}$ of $USp(8)$ decomposes as ${\bf 27} + {\bf 1}$. Again one
can check that all gauge and gravitational anomalies cancel.

\subsubsection{T-duality and the ${\bf USp(32)}$ string theory}

We conclude this section by commenting briefly on an
interesting point, concerning the T-duals of the models we have just
constructed. Notice that the action of T-duality on anti-D-branes is the
same as for D-branes. Therefore, if we perform a T-duality along the four
compact directions in the orientifolds of the previous subsections, we
obtain a set of models with 32 \Dn-branes and 32 D5-branes, giving
rise to the spectra we have presented.

It seems striking to encounter consistent models with 32 \Dn-branes,
since in usual type IIB orientifolds, cancellation of untwisted tadpoles
imposes the presence of either 32 D9-branes (if $\Omega$ belongs to the
orientifold group) or no D9-branes (if it does not). However, reference
\cite{sugimoto} discussed the existence of an $\Omega$ projection of
ten-dimensional type IIB string theory, such that the RR  charge of the
corresponding crosscaps (that is, the charge of the O9-plane) is opposite
to the usual one. Cancellation of tadpoles in this model requires the
introduction of $-32$ units of D9-brane charge, which can be minimally
accomplished by introducing 32 \Dn-branes. This construction leads
to a ten-dimensional tachyon-free non-supersymmetric string theory of open
and closed unoriented strings, with $USp(32)$ gauge group. Clearly, the
T-duals of the models we have constructed in the previous section
correspond to compactifications (with vector structure) of this
ten-dimensional non-supersymmetric on $T^4/\IZ_N$ spaces. In other words,
they correspond to modding out type  IIB theory on  $T^4/\IZ_N$, by the
orientifold action $\Omega$ introduced in \cite{sugimoto}.

The T-duality relation between the type IIB orientifolds of the previous
subsections and the type IIB orientifolds with the $\Omega$ projection in
\cite{sugimoto} seems to suggest interesting relations between the
$SO(32)$ and $USp(32)$ theories, at least, after compactification. It
would be important to gain a better understanding of such relations.
Certainly, the construction of further examples of six-dimensional
consistent compactifications of both theories will provide new insights
into these issues. We hope our techniques are useful in these
investigations.

\section{Four-dimensional models}

In this section we present some examples of four-dimensional type IIB
orientifolds containing branes and antibranes. The models are tachyon-free
and give rise to chiral spectra. The class of models that can be
constructed is presumably very large. The examples we present have been
selected to illustrate the following interesting possibility, which has
direct application in phenomenological model building. It is possible to
construct explicit type IIB orientifolds where $\NN=1$ chiral
supersymmetric sectors of D-branes are spatially separated (in the
internal space) from non-supersymmetric sectors of anti-D-branes. This
type of models provides an explicit realization of the supersymmetry
breaking scenario where the standard model is embedded in a set of
supersymmetric branes, whereas supersymmetry is broken (in our case, at
the string scale) in a hidden sector of antibranes. This scenario could be
phenomenologically viable for suitable choices of the string and
compactification scales, and fits nicely into the circle of ideas recently
developed about string theory vacua with large or largish dimensions
\cite{antoniadis,larged,bbiq,rs}.

\subsection{A $\IZ_6'$ model with non-supersymmetric and supersymmetric
sectors}

Here we  describe a chiral tachyon free four-dimensional example
with $\NN=1$ supersymmetric sectors, and non-supersymmetric `hidden'
sectors. Branes and antibranes are stuck at different fixed points in the
compact space and are unable to move off into the bulk. Therefore, the
model is stable against the kind of annihilation and decay mentioned in
section 3.1.

The model is based on the $\IZ_6'$ orientifold, with twist $v=(1,-3,2)/6$.
We will make use of the property, discussed in \cite{afiv, abiu}, that
tadpoles can be satisfied placing D5$_3$-branes (and in our case, also
\Df$_3$-branes) at different fixed points in the second complex
plane.

Let us consider the following Chan-Paton matrix for the D9-branes
\beqa
\gamma_{\theta,9}=\diag(e^{\pi i\frac 16}1_4, e^{\pi i\frac 36}1_8,
e^{\pi i\frac 56}1_4, e^{\pi i\frac 76}1_4, e^{\pi i\frac 96}1_8, e^{\pi
i\frac{11}{6}}1_4)
\eeqa
Let us also place some D5-branes at the origin in the second complex
plane, with Chan-Paton factors
\beqa
\gamma_{\theta,5,0}=\diag(e^{\pi i\frac 16}1_4, e^{\pi i\frac 56}1_4,
e^{\pi
i\frac 76}1_4, e^{\pi i\frac{11}{6}}1_4)
\eeqa
We place \Df-branes at another fixed point, labeled `1', with
matrices
\beqa
\gamma_{\theta,\bar{5},1}=\diag(e^{\pi i\frac 36}1_R,e^{\pi i\frac 96}1_R)
\eeqa
and some D5-branes at yet another fixed point, labeled `2', with matrices
\beqa
\gamma_{\theta,5,2}=\diag(e^{\pi i\frac 36}1_{8+R},e^{\pi i\frac
96}1_{8+R})
\eeqa
The matrices $\gamma_{\Omega}$ are defined following the conventions in
section 2.1.

The choice of matrices above is just a small variation of the  model in
\cite{afiv}, and satisfies the tadpole conditions. The spectrum in
the 99 sector is the following $\NN=1$ theory
\beqa
& U(4)\times U(8)\times U(4) & \nonumber \\
& (6,1,1)+(4,{\ov 8},1)+(1,8,{\ov 4})+(1,1,6) + ({\ov 4},{\ov 8},1)+
&\nonumber \\
&+(4,1,{\ov 4}) +(1,8,4) + (4,1,4)+({\ov 4},1,{\ov 4})+(1,28,1)+ (1,{\ov
{28}},1)&
\eeqa
The D5-branes at the origin form a $\NN=1$ supersymmetric sector. For
instance, their 55 strings gives the following $\NN=1$ supersymmetric
spectrum
\beqa
& U(4)\times U(4) & \nonumber\\
& (6,1)+(1,6)+(4,{\ov 4})+ (4,4)+({\ov 4},{\ov 4})&
\eeqa
The corresponding 59 matter is given by the $\NN=1$ chiral multiplets
\beqa
(4,1;4,1,1)+(1,4;1,{\ov 8},1)+(1,{\ov 4};1,1,{\ov 4})+({\ov 4},1;1,8,1)
\eeqa
The D5-branes at the fixed point labeled `2' also give an $\NN=1$ sector,
as follows
\beqa
\begin{array}{cc}
& U(8+R) \\
 55 \quad & \Yasymm + \bYasymm  \\
 59 \quad & \quad (\fund;{\ov 4},1,1) + (\antifund;1,1,4)
\end{array}
\eeqa
Finally, the \Df-branes give a gauge group $U(R)$ with the
following {\em non-supersymmetric} matter content
\beqa
\begin{array}{cccc}
{\rm {\bf Sector}}& {\rm {\bf Complex\;\, Scalars}}& {\rm {\bf
Fermion}}_+& {\rm {\bf Fermion}}_-\\
\bar{5}\bar{5} & \Yasymm+\bYasymm & \Ysymm +\bYsymm &  {\rm Adj.}\\
\bar{5}9+9\bar{5} & (R;4,1,1)+({\ov R};1,1,{\ov 4}) & & (R;{\ov4},1,1)
+ ({\ov R};1,1,4)
\end{array}
\eeqa
It is easy to check that non-abelian anomalies cancel. Also, $U(1)$
anomalies should cancelled by the GS mechanism proposed in \cite{iru1}.

This model provides a specific realization of a scenario where
supersymmetry is broken explicitly at the string scale in an anti-D-brane
hidden sector, distant from the $\NN=1$ chiral supersymmetric sector. We
expect this model to belong to a more general class of explicit type IIB
orientifold realizing this type of scenario. Hopefully, this class of
models may include theories with more realistic spectra.

In the  model we have constructed, if the standard model would be
embedded in one of the supersymmetric D5-brane sectors, supersymmetry
breaking would be felt through loops involving states in the 99 and
$9\bar{5}$ sectors. Therefore, supersymmetry breaking is gauge-mediated.
It should not be particularly difficult to construct examples where the
supersymmetry breaking is gravity-mediated. This could be accomplished,
for instance, by introducing Wilson lines that project all 59 states
in the visible sector out.

As mentioned above, branes and antibranes in the model are stuck at fixed
points in the compact space, and therefore cannot move off to the bulk and
lead to annihilation\footnote{Upon a closer look, a possible process of
anihilation may take place in the region of moduli space where the points
labeled `1' and `2' are very close, and a winding mode stretched between
branes and antibranes becomes tachyonic. Even though the  anihilation
would involve objects at different fixed points, it seems to be consistent with
tadpole cancellation in the final state. This new kind of instability is
not generic and is absent in other models, in particular the $\IZ_3$
orientifold in the next section.}. However, there exist non-cancelled forces
between the branes and antibranes, whose strength depends on the distances
between the corresponding fixed points. Another interesting feature of
this models is that by increasing $R$ the gauge groups in the model can
be made as large as desired, with no apparent inconsistency. The only
prize to pay would be an increase in the vacuum energy of the model
(cosmological constant) since, roughly speaking,  $R$ controls the number
of excess brane-antibrane pairs. In our final remarks we speculate
on the possible role these two dynamical issues may play in moduli
stabilization.

\subsection{A $\IZ_3$ example with non-supersymmetric  and supersymmetric
sectors}

It is easy to construct further examples of four-dimensional models with
supersymmetric and non-supersymmetric sectors. Here we discuss a model based on the $\IZ_3$ orientifold \cite{ang}. In order to
cancel the untwisted tadpole the orientifold  action requires the
introduction of 32 D9-branes and zero net number of D5-branes. However,
one can introduce an equal number of D5-branes and \Df-branes
(wrapping, for instance, the third complex plane) and
obtain new (non-supersymmetric) consistent models. The tadpole
cancellation conditions read
\beqa
\Tr \gamma_{\theta,9} + 3(\Tr\gamma_{\theta,5,L}-
\Tr\gamma_{\theta,\bar{5},L}) = -4
\label{tadpz3}
\eeqa
for any of the nine fixed points in the two first complex planes. Let us
consider a particular case, with
\beqa
\gamma_{\theta,9}=\diag(1_{16},e^{2\pi i\frac 13} 1_8,e^{2\pi i \frac 23}
1_8)
\eeqa
that is, $\Tr \gamma_{\theta,9}=8$. This choice implies that no fixed
point can be completely empty. Tachyons are most easily avoided by
considering the fixed points to contain only D5-branes or only
\Df-branes. We label such fixed points by indices $M$, $K$,
respectively. Twisted tadpoles (\ref{tadpz3}) require $\Tr
\gamma_{\theta,5,M}=-4$, $\Tr\gamma_{\theta,\bar{5},K}=4$, therefore
\beqa
\gamma_{\theta,5,M} & = & \diag(1_{n_M}, e^{2\pi i\frac 13}1_{n_M +4},
e^{2\pi i\frac 23} 1_{n_M +4}) \nonumber \\
\gamma_{\theta,\bar{5},K} &=&\diag(1_{m_K +4}, e^{2\pi i\frac 13}1_{m_K},
e^{2\pi i\frac 23} 1_{m_K}) \nonumber \\
\eeqa
Since no net D5-brane charge is allowed
\beqa
\sum_M (3n_M +8)=\sum_K (3m_K +4)
\eeqa
A particular solution is obtained by choosing $n_M=0$, $m_K=0$, and
three fixed points containing only D5-branes, and six with only
\Df-branes. This choice has the virtue that all branes and
antibranes are stuck at the fixed points.

The spectrum in the general case can be computed using our rules. Since it
is not particularly interesting, we spare the reader these details. Let us
just mention that the gauge group of the model is
\beqa
SO(16)\times U(8) \times \prod_M [\, USp(n_M)\times U(n_M+4)\,] \times
\prod_K [\,USp(m_K +4)\times U(m_K) \,] \quad
\eeqa
and the spectrum of fermions is
\beqa
\begin{array}{ccc}
{\bf 99} & {\rm Fermions_+} & 3(16,\ov{8}) + 3(1,28) \\
         & {\rm Fermions_-} & (\Yasymm,1) + (1,{\rm Adj}) \\
{\bf 5_M9} & {\rm Fermions_+} & (n_M,1;1,\ov{8}) + (1,n_M+4;1,8) +
(1,{\ov {n_M+4}};16,1)  \\
{\bf \bar{5}_K9} & {\rm Fermions_-} & (m_K+4,1;1,\ov{8}) + (1,m_K;1,8) +
(1,{\ov {m_K}};16,1) \\
{\bf 5_M5_M} & {\rm Fermions_+} & 3(n_M,{\ov{n_M+4}}) + 2(1,\Yasymm)
+(1,\Ysymm)\\
             & {\rm Fermions_-} & (\Ysymm,1) + (1,{\rm Adj}) \\
{\bf \bar{5}_K\bar{5}_K} & {\rm Fermions_+} & 3(m_K+4,{\ov{m_K}}) +
2(1,\Ysymm) +(1,\Yasymm) \\
 & {\rm Fermion_-} & (\Yasymm,1)+ (1, {\rm Adj})
\end{array}
\eeqa
One can check that all anomalies cancel (to show that, one has to use the
fact that none of the nine fixed points in the first two complex planes is
empty).

Therefore we obtain a set of models with diverse numbers of supersymmetric
and non-supersymmetric sectors. Most of the comments in the previous
section apply to these models as well, most notably the gauge-mediated
nature of supersymmetry breaking in the supersymmetric $D5$-brane sectors.

We hope the examples in this Section suffice to illustrate the
construction of  consistent four-dimensional orientifolds with
antibranes, and the relative ease with which
supersymmetric and non-supersymmetric sectors within the same
construction can be implemented

\section{Conclusions}

In this paper we have provided the basic rules  for the construction of
type IIB orientifold models with branes and antibranes. The models
present the attractive feature that closed string tachyons are
automatically absent, since supersymmetry is preserved in the closed
string sector. Moreover, open string tachyons from strings stretching
between branes and antibranes can be avoided. Therefore many new
tachyon-free non-supersymmetric models can be constructed using
this simple idea. We have provided several explicit examples in six and
four dimensions, which illustrate the generic features of these models.

A particular interesting application of these constructions, from the
phenomenological viewpoint, is that they allow the presence of $\NN=1$
supersymmetric sectors of branes and supersymmetry breaking sectors of
antibranes, spatially separated in the compact space.

The models may also play an interesting role in the web of string dualities
involving other non-supersymmetric strings. We have discussed that
type IIB orientifolds on $T^4/\IZ_N$ with vector structure are related by
T-dualtity to analogous compactifications of the non-supersymmetric
$USp(32)$ theory of \cite{sugimoto}. We hope further research on these
vacua will uncover new interesting properties.

Finally, we would like to point out two important issues we have not
addressed, concerning the dynamics of moduli of these models. As briefly
mentioned, branes and antibranes in the model suffer attractive forces
due to the lack of supersymmetry. If these objects are stuck at different
fixed points, their attraction presumably leads to a potential for the
compactification moduli, pushing their vevs to small radii. If the models
contain other sources of non-trivial moduli dynamics, pushing the radii to
large values, this effect would contribute to their stabilization.  The
second point refers to  the cosmological constant. Computing the
vacuum energy by simply taking into account the tension of branes,
antibranes and orientifold planes ({\em i.e.} ignoring for the moment
interaction energies), it is easy to see that the models we have
discussed have a positive cosmological constant, which is, roughly,
inversely proportional to the string coupling constant. Even without
having addressed its specific value, which furthermore would be rather
model-dependent, it is tempting to interpret this  vacuum energy as a
potential for the dilaton. From that point of view, the model would force
the string coupling to become strong. If the model has sources of
dilaton potential with a runaway behaviour to small coupling (for
instance, a gaugino condensate), the dilaton would be stabilized.

In any case, interesting dynamics for the moduli
seems to arise from the absence of supersymmetry. Whether stabilization takes
place, and whether the resulting values for the compactification scale
and string coupling are of phenomenological interest, depends on dynamical
information we have not yet studied. We hope further reseach in these
models succeeds in uncovering their rich dynamics and their
phenomenological potential.

\bigskip

\centerline{\bf Acknowledgements}
We are grateful to  A.~Karch,
J.~Park, and especially L.~E.~Ib\'a\~nez, for useful discussions.
G.A. thanks colleagues at the Extended Workshop in String Theory,
AS-ICTP, for stimulating discussions. A.~M.~U. thanks M.~Gonz\'alez for
kind encouragement and support. G.A work is partially supported by APCyT
grant 03-03403. The work of A.~M.~U. is supported by the Ram\'on Areces
Foundation (Spain).

\end{document}